\documentstyle[twocolumn,aps]{revtex}
\input epsf
\begin{document}
\draft
\twocolumn[\hsize\textwidth\columnwidth\hsize\csname@twocolumnfalse\endcsname

\title{Drift-Controlled Anomalous Diffusion: A Solvable Gaussian Model}
\author{Fabrizio Lillo and Rosario N. Mantegna}
\address{
Istituto Nazionale per la Fisica della Materia, Unit\`a di Palermo\\
and\\ Dipartimento di Energetica ed Applicazioni di Fisica,
Universit\`a di Palermo, Viale delle Scienze, I-90128, Palermo, Italia}
\maketitle
\begin{abstract}

We introduce a Langevin equation characterized by a time dependent 
drift. By assuming a temporal power-law dependence of the drift 
we show that a great variety of behavior is observed in the 
dynamics of the variance of the process. In particular diffusive, 
subdiffusive, superdiffusive and stretched exponentially diffusive 
processes are described by this model for  
specific values of the two control parameters.
The model is also investigated in the presence of an external 
harmonic potential. We prove that the relaxation to the
stationary solution is power-law in time with an exponent 
controlled by one of model parameters.

\end{abstract}
\vskip2pc]


Diffusive stochastic processes, i.e. stochastic processes $x(t)$
characterized by a linear grows in time of the variance 
$<x^2(t)> \propto t$ are quite common in physical systems. However
deviations from a diffusive process are observed in several 
stochastic systems.
Superdiffusive ($<x^2(t)> \propto t^\nu$ with $\nu>1$) and
subdiffusive ($<x^2(t)> \propto t^\nu$ with $\nu<1$) random
processes have been detected and investigated in physical and
complex systems. A classical example of superdiffusive random process
is Richardson's observation that two particles moving
in a turbulent fluid which at time $t=0$ are originally placed 
very close the 
one with the other have a relative separation 
$\ell$ at time $t$ that follows 
the relation $<\ell^2(t)> \propto t^3$ \cite{Richardson26}. 
Most recent examples include anomalous kinetics in chaotic dynamics 
due to flights and trapping \cite{Geisel85,Shlesinger93}, anomalous 
diffusion in aggregate of amphiphilic molecules \cite{Ott90} and 
anomalous diffusion in a two-dimensional rotating flow 
\cite{Solomon93}. Subdiffusive stochastic processes
have also been detected and investigated. Examples includes
charge transport in amorphous semiconductors \cite{Scher75,Gu96}
and the dynamics of a bead in polymers \cite{Amblard96}.
Another class of stochastic processes which are not diffusive 
in a simple way is the one characterized by a variance with a
stretched exponential time dependence. When a such process is
Gaussian distributed the probability of return to the origin
$P_0 (t)$ is described by the Kohlrausch law  
$P_0 (t) \propto \exp [-t^\nu]$ with $\nu<1$. Similar behaviors are observed
in glassy systems and in random walks in ultrametric
spaces \cite{Ogielski85}.

The modeling of some of the above discussed anomalous diffusing 
stochastic processes has been done by using a variety of approaches. 
To cite some examples, we recall that superdiffusive and subdiffusive 
processes have been modeled by writing down a generalized 
diffusion equation 
\cite{Richardson26,Batchelor52,Hentschel84}, by introducing 
L\'evy walks models \cite{Shlesinger82}, by using a fractional 
Fokker-Planck equation approach \cite{Metzler99} or by using 
``ad hoc" stochastic models such as, for example, the fractional 
Brownian motion \cite{Mandelbrot68}. 

In this rapid communication we introduce a class of Langevin equations
able to describe all the different anomalous 
regimes discussed above for Gaussian processes.  
Specifically, we study the properties of the class of Langevin 
equations
\begin{equation}
\dot x+\gamma(t)x=\Gamma(t),
\end{equation}
where $\gamma(t)$ is a function of time $t$ and $\Gamma(t)$ is a 
Langevin force with zero mean and with a correlation function given by
$<\Gamma(t_2)\Gamma(t_1)>=D\delta(t_2-t_1)$. Equation (1) describes an 
Ornstein-Uhlenbeck process \cite{Uhlenbeck30} in the particular 
case of $\gamma(t)=\gamma$. This 
equation is linear and solvable. For the sake of simplicity 
we set the boundary condition of Eq. (1) at $t=0$.  
The formal solution of Eq. (1) is
\begin{equation}
x(t)=x(0)G(t)+G(t)\int_0^t\frac{\Gamma(s)}{G(s)}ds,
\end{equation}
where $G(t) \equiv \exp [ -\int_0^t \gamma(s)ds ]$.
By using this formal solution and all order correlation functions 
of $\Gamma(t)$ we obtain all central moments of $x(t)$. The first two 
central moments are given by
\begin{eqnarray}
&&<x(t)>=x(0)G(t)\equiv \mu(t), \nonumber \\
&&<(x(t)-\mu(t))^2>=DG^2(t)\int_0^t\frac{1}{G^2(s)}ds\equiv \sigma^2(t).
\end{eqnarray}
The general relation between higher-order even central moments 
and the second central moment of the investigated processes is 
the one observed in a Gaussian process. Moreover odd central moments 
are zero hence we conclude that the stochastic processes described by 
Eq. (1) is Gaussian.

We now consider the two-time correlation functions of the 
process $x(t)$ and of its time derivative $\dot x(t)$. 
In the following we label the two times $t_1$ and $t_2$
of the correlation functions in such a way that 
$t_2 \geq t_1$. By using the formal solution of Eq. (2) 
we determine the two-time correlation function for the 
random variable $x(t)$
\begin{equation}
<x(t_1)x(t_2)>=\mu(t_1)\mu(t_2)+\frac{G(t_2)}{G(t_1)}\sigma^2(t_1).
\end{equation}
In general, the correlation function $<x(t_1)x(t_2)>$ is 
not a function of $t_2-t_1$ and therefore the process is 
usually non-stationary.

By starting from the correlation function of $x(t)$ and 
from the formal solution of 
the Langevin equation we obtain the two-time correlation 
function of $\dot x(t)$ as
\begin{eqnarray}
<\dot x(t_1) \dot x(t_2)>=\mu_v(t_1)\mu_v(t_2)+D\delta(t_2-t_1)
\nonumber \\
+\gamma(t_2) G(t_2) F(t_1),
\end{eqnarray}
where $\mu_v(t)=-\gamma(t) \mu(t)$ indicates the mean of the time derivative 
$\dot x(t)$ and $F(t_1) \equiv \left[\gamma(t_1)
\sigma^2(t_1)-D \right]/G(t_1)$. The two-time correlation 
function of $\dot x(t)$ 
is the sum of a delta function and of a smooth function.

The Fokker-Planck equation associated to the Langevin equation 
given in Eq. (1) is
\begin{equation}
\frac{\partial \rho}{\partial t}=\frac{\partial}{\partial x}
\left( \gamma(t) x \rho\right) 
+\frac{D}{2} \frac{\partial^2 \rho}{\partial x^2}.
\end{equation}
This Fokker-Planck equation is the same as the Smoluchowski 
equation of a Brownian particle moving in a harmonic oscillator 
with a time dependent potential $U(x)\propto x^2\gamma(t)$.
In our study we consider both positive and negative 
values of $\gamma(t)$. For positive values of $\gamma(t)$ 
the position $x=0$ is a stable equilibrium position whereas 
in the opposite case $x=0$ is an unstable equilibrium position.   

We calculate the two-time conditional probability density 
$P(x_2,t_2|x_1,t_1)$ as the Green function of the Fokker-Planck equation. 
In our formalism $t_1\le t_2$. Our determination is done by 
working with the Fourier transform of $P(x_2,t_2|x_1,t_1)$ with 
respect to the $x$ variable. The equation for the Fourier 
transform of $P(x_2,t_2|x_1,t_1)$ is a first order partial 
differential equation, which can be solved by the methods
of characteristics. We obtain
\begin{eqnarray}
P(x_2,t_2|x_1,t_1)=\frac{1}{\sqrt{2\pi s^2(t_2,t_1)}} \nonumber \\
\exp\left(-\frac{(x_2-m(t_2,t_1)x_1)^2}{2s^2(t_2,t_1)}\right),
\end{eqnarray}
where $m(t_2,t_1)=\exp \left[-\int_{t_1}^{t_2}\gamma(y)dy\right]$,
and $s^2(t_2,t_1)= D \int_{t_1}^{t_2}\exp \left[-2 \int_{z}^{t_2}
\gamma(y)dy\right] dz$.
Hence the transition probability of Eq. (7) is a Gaussian 
transition probability. Moreover, Eq. (7) satisfies the 
Chapman-Kolmogorov equation. In fact from a direct 
integration one can verify that 
$P(x_3,t_3|x_1,t_1)=\int P(x_3,t_3|x_2,t_2)P(x_2,t_2|x_1,t_1)dx_2$.

In the rest of this rapid communication we restrict our attention to 
the class of Langevin equations 
with a drift term which has a temporal behavior 
of the form
\begin{equation}
\gamma(t)\sim a / t^{\beta}
\end{equation}
for large time values.
We study the stochastic process of Eq. (1) for different values 
of parameters $a$ and $\beta$. Specifically we focus on the asymptotic 
temporal evolution of the variance and of the two-time correlation 
function of $\dot x(t)$. We recall that for $a=0$ Eq. (1) describes 
a Wiener process with a variance increasing in a diffusive way, 
$\sigma^2(t) \sim t$, and a delta correlated $\dot x(t)$. 
When $\beta=0$ Eq. (1) describes an Ornstein-Uhlenbeck process 
and two regimes are observed depending on the sign of $a$. 
When $a>0$ the stochastic process has a stationary Gaussian solution, 
whereas when $a<0$ there is no stationary state and the variance 
increases asymptotically in an exponential way, 
$\sigma(t)\sim \exp(2|a| t)$ \cite{Uhlenbeck30,Risken}. 
The two-time correlation of the velocity decreases in an 
exponential way as $\exp(-|a|(t_2-t_1))$.  

The cases considered above are known, in addition to these cases we 
observe a large variety of new behaviors controlled by the specific 
values of parameters $a$ and $\beta$. 
By investigating the $(\beta,a)$ set of parameters, we detect 
different anomalous behavior that we discuss below systematically 
by considering different regions of the $\beta$ parameter. 

(i) Region with $\beta > 1$. The process $x(t)$ is diffusive and its 
variance increases linearly with time for any value of $a$. 
The two-time correlation function of $\dot x(t)$ can be obtained
starting from Eq. (5). A direct calculation gives
\begin{equation}
<\dot x(t_1) \dot x(t_2)>\sim \frac{a}{t_2^{\beta}} 
\exp \left(\frac{-a t_2^{1-\beta}}{1-\beta }\right)F(t_1). 
\end{equation}
The process 
$\dot x(t)$ can be positively or negatively 
correlated depending on the sign of $a$. When $a>0$ $(a<0)$ the 
correlation is negative (positive). This property is valid for any
value of $\beta$ parameter. By investigating the explicit form 
of Eq. (9) one observes that the correlation function decreases 
as a function of $t_2$ with a power-law 
dependence, $<\dot x(t_1) \dot x(t_2)>\sim 1/t_2^{\beta}$.  

(ii) Region with $\beta=1$. In this case we observe two regimes. When $a>-1/2$ 
the variance increases in a diffusive way $\sigma^2(t) \sim t$. 
We find a different behavior when $a<-1/2$. In fact, by using 
Eq. (3) one can show that 
\begin{equation}
\sigma^2(t) \sim t^{2|a|}.
\end{equation}
Therefore the particle performs a Gaussian superdiffusive 
random process. At the boundary value $a=-1/2$ the variance increases 
in a log divergent way as $\sigma^2(t) \sim t\log t$. 
The two-time correlation function of $\dot x(t)$ is determined 
starting from Eq. (5). An explicit calculation gives 
\begin{equation}
<\dot x(t_1) \dot x(t_2)>\sim \frac{a}{t_2^{1+a}}F(t_1). 
\end{equation}
The two-time correlation function of $\dot x(t)$ shows a power-law 
time dependence and the $\dot x(t)$ process is a strongly dependent
random process \cite{Cassandro78}. We wish to point out that
when $a=-1$ the diffusion of the $x(t)$ process is ballistic.
This specific case has been already investigated by E. Nelson
in the framework of stochastic mechanics. 
The Ito equation describing the stochastic process associated with 
the free evolution of a Gaussian quantum wave packet is \cite{Nelson67} 
\begin{equation}
d x(t)=\frac{t-c}{t^2+c^2}x~dt+dw(t),
\end{equation}
where $w(t)$ is a Wiener process and $c$ is a constant .
This stochastic equation describes the same random process
of Eq. (1) for large values of $t$. 

(iii) In the region $0<\beta<1$ we observe two regimes, which 
depend on the sign of $a$. When $a>0$ the variance increases as
\begin{equation}
\sigma^2(t) \sim t^{\beta}.
\end{equation}
This behavior is the customary behavior observed in subdiffusive 
random process. The two-time correlation function of $\dot x(t)$ behaves 
asymptotically as
\begin{equation}
<\dot x(t_1) \dot x(t_2)>\sim -\frac{aD}{t_2^{\beta}} \exp \left( -\frac{a}{1-\beta}(t_2^{1-\beta}-t_1^{1-\beta})\right).
\end{equation} 
If the time interval $\tau\equiv t_2-t_1$ is shorter than $t_1$ 
\cite{Keshner82} the power-law term dominates in this equation 
and the stochastic process is power-law anti-correlated. For 
$\tau >> t_1$ the two-time correlation function decreases
exponentially.  

When $a<0$ the variance increases as a stretched exponential
\begin{equation}
\sigma^2(t) \sim \exp \left[\frac{2|a|}{1-\beta}t^{1-\beta}\right].
\end{equation}
Since the process is Gaussian, the probability of return to 
the origin follows the Kohlrausch law, 
$\rho(x(0),t)= 1/\sqrt{2 \pi} \sigma(t) \sim \exp{\left[-t^{1-\beta}\right]}$. 
This kind of anomalous diffusion has been observed in 
glasses and in random walks on an ultrametric space \cite{Ogielski85}. 
The two-time correlation function of $\dot x(t)$ increases with time 
as Eq. (14).

(iv) Region $\beta <0$. This region is essentially different from 
the previous ones because the absolute value of the drift term 
increases in time and eventually diverges.  In this case we also observe 
two regimes depending on the sign of $a$. When $a<0$ the time
evolution of the variance is formally the same as Eq. (15) 
of case (iii). In this region of $\beta$ parameter the variance
increases  more than exponentially in time.
When $a>0$ we find that the variance decreases with time with 
a power-law dependence $\sigma^2(t) \sim 1/t^{|\beta|}$.
By using the Smoluchowski picture, we can interpret this behavior 
as the motion of Brownian particle moving in a time dependent
potential which leads to a localization of the particle in the 
point $x=0$. 
For both regimes the two-time correlation function of 
$\dot x(t)$ is given by Eq. (9). 

We summarize the above discussed variety of diffusive behavior 
of $\sigma^2 (t)$ in Table I. 

The results obtained above refer to $x(t)$ random processes which 
are not stationary. We now consider the problem 
of a process $x(t)$ whose dynamics is controlled by a
modified version of Eq. (1) in which the effects of the presence
of an ``external" time-independent potential are taken into account.  
To this end we consider the specific case of a overdamped 
particle of mass M moving in a viscous medium in the presence 
of a potential having a time-dependence of the kind of Eq. (8) and a 
time-independent part. The equation of motion of such a system is
\begin{equation}
M \eta \dot x + g(t) x - F(x) =M \tilde \Gamma(t), 
\end{equation}
where $\eta$ is the friction constant and $\tilde \Gamma(t)$ is 
the Langevin force with diffusion constant $2\eta k_B T/M$.
This equation is formally equivalent to
\begin{equation}
\dot x+\gamma(t)x+V'(x)=\Gamma(t),
\end{equation}
when $V'(x)=-F(x)/M \eta$, $\gamma(t)=g(t)/M\eta$ and 
$D=2 k_B T/ M \eta$. The prime in $V(x)$ indicates spatial 
derivative. It is worth pointing out that when $\gamma(t)$ 
goes to zero as $t$ increases (as, for example, in the case 
$\gamma(t)\sim a/t^{\beta}$ with $\beta>0$) Eq. (17)
might have a stationary solution. The presence of a 
stationary solution depends on the exact shape of $V(x)$.

To investigate in a concrete example the relaxation 
dynamics of the probability density function of $x(t)$ 
towards the stationary solution we study Eq. (17) in the
presence of an external harmonic potential, $V(x)=\frac{1}{2}kx^2$. 
In this case the process has a stationary state. 
A general solution of Eq. (17) is found by using the 
substitution $\gamma(t) \rightarrow \gamma(t)+k/M\eta$ 
in Eq. (2). 
In this case the variance of the process is equal to
\begin{equation}
\sigma^2(t)=De^{-2kt/M\eta}G^2(t)\int_0^t\frac{e^{2ks/M\eta}}{G^2(s)}ds.
\end{equation}
The asymptotic stationary value of $\sigma^2(t)$ is 
$\sigma^2_{st}\equiv D M\eta/2k=k_BT/k$ which is independent 
from the parameters $a$ and $\beta$.
However, we observe a relaxation dynamics whose functional form 
is controlled by the values of $a$ and $\beta$ parameters. 
To detect the different relaxation dynamics we evaluate numerically 
the integral in Eq. (18) by setting $\gamma(t)=a/(\tau^\beta+t^{\beta})$. 
In Fig. 1 we show in a log-log plot the quantity 
$\Delta(t)\equiv|\sigma^2(t)- \sigma^2_{st}|/\sigma^2_{st}$ as 
a function of time. The quantity $\Delta(t)$ provides a measure
of the distance of the system from the stationary behavior. 
In Fig. 1 we show that the quantity $\Delta(t)$ decreases 
following the power-law behavior $\Delta(t)\propto 1/t^{\beta}$  
for large values of $t$ and  for all the investigated values 
of the parameters $a$ and $\beta$. In particular, when $a>0$, 
$\sigma^2(t)- \sigma^2_{st}$ goes to zero as a negative value
whereas when $a<0$ the same quantity goes to zero 
taking positive values. In order to 
illustrate this result we show $\sigma^2(t)$ as a function of 
time when $\beta =0.6$ and $a=\pm 1$ in the inset of Fig. 1. 
For $0<\beta \leq 1$ (therefore including 
subdiffusive, superdiffusive and stretched exponentially
diffusing processes) it is not possible to define a 
characteristic time scale for the convergence of 
$\sigma^2(t)$ during the process of relaxation. This is due to
the fact that the integral of $\int_{t_1}^{\infty}
\Delta(t)/\Delta(t_1) dt = \infty$. Although 
a power-law behavior is still observed when $\beta>1$, it is 
worth pointing out that in this interval of $\beta$ parameter 
a typical time scale might be determined by considering
the above discussed integral which is finite in this region of
$\beta$ parameter.

In conclusion the Langevin equations (1) and (17) 
with the choice of Eq. (8)
describe non-stationary and stationary random processes showing a
wide class of (normal and anomalous) diffusion. When a stationary
state exists, the relaxation dynamics to the stationary state 
has a power-law time dependence. The processes modeled by Eqs
(1) and (17) are characterized by a time dependent drift term 
in the associated Fokker-Planck equation. Our model is 
complementary to the Batchelor's description of anomalous 
diffusion obtained by assuming a time-dependent diffusion 
term \cite{Batchelor52}. 
Equations (1) and (17) can be used to model metastable 
systems in which one of the physical observables,
such as, for example, the viscosity, is time dependent. 
They can also be used to develop simple and efficient 
algorithms generating realizations of random processes 
with controlled anomalous diffusion. 

The authors thank INFM and MURST for financial support. F. Lillo 
acknowledges FSE-INFM for his fellowship.



\begin{figure}[c]
\epsfxsize=3.1in
\epsfbox{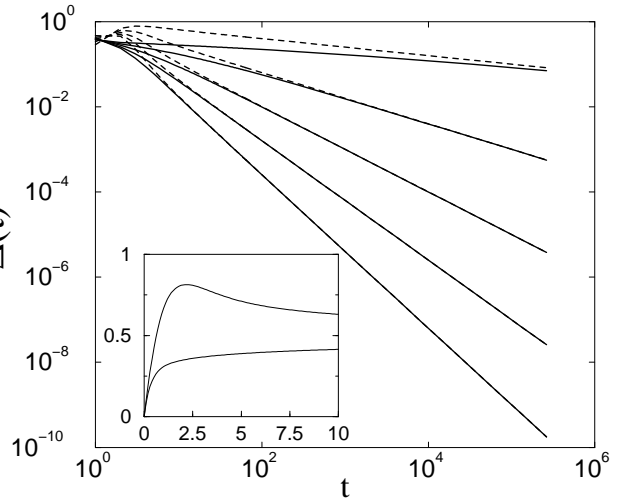}
\caption{Theoretical estimation of the normalized 
difference of the variance at time $t$ from the 
stationary value $\sigma^2_{st}$, $\Delta(t)$ as a function of 
time. Different curves refer to different values of
the control parameters $a$ and $\beta$. The $a$ values 
are $a=1$ (solid lines) and $a=-1$ (dashed lines).
The parameter $\beta$ assumes the values 0.2, 0.6, 
1.0, 1.4 and 1.8 from top to bottom. In the inset 
we show a typical time evolution of $\sigma^2(t)$ 
obtained by setting $\beta=0.6$ and two different 
values of $a$, $a=1$ (bottom curve) and $a=-1$ 
(top curve). The other parameters are $\tau=1$,
$k/M\eta=1$ and $D=1$}
\label{fig1}
 \end{figure}

\begin{table}
\caption{Summary of the different diffusion regimes. 
The constant $C\equiv2|a|/(1-\beta)$}
\begin{tabular}{cccc}
$\beta$ & $a$ & $\sigma^2(t)$ & Description  \\
\tableline
\tableline
$\beta$ & $a=0$ & $t$ & Wiener (diffusive)\\
\tableline
$\beta>1$ & $a>0$ & $t$ & diffusive  \\
$\beta>1$ & $a<0$ & $t$ & diffusive  \\
\tableline
$\beta=1$ & $a>-1/2$ & $t$ & diffusive  \\
$\beta=1$ & $a=-1/2$ & $t\log t$ & log divergent \\
$\beta=1$ & $a<-1/2$ & $t^{2|a|}$ & superdiffusive  \\
\tableline
$0<\beta<1$ & $a>0$ & $t^{\beta}$ & subdiffusive  \\
$0<\beta<1$ & $a<0$ & $\exp[C t^{1-\beta}]$ & 
less than \\
~ & ~ & ~ & exponentially diffusive  \\
\tableline
$\beta=0$ & $a>0$ & $1-\exp (-2at)$ & Ornstein-Uhlenbeck  \\
$\beta=0$ & $a<0$ & $\exp(2|a|t)$ & 
exponentially diffusive  \\
\tableline
$\beta<0$ & $a>0$ & $1/t^{|\beta|}$ & localized  \\
$\beta<0$ & $a<0$ & $\exp[C t^{1-\beta}]$ & 
more than \\
~ & ~ & ~ & exponentially diffusive  \\
\end{tabular}
\end{table}
\end{document}